\documentstyle[prb,preprint,aps]{revtex}    

\begin{document}
\draft
\newcommand{\beq}{\begin{equation}}
\newcommand{\eeq}{\end{equation}}
\newcommand{\beqa}{\begin{eqnarray}}
\newcommand{\eeqa}{\end{eqnarray}}

\title{Supercurrent flow through an
effective double barrier structure}

\author{Ivar Zapata and Fernando Sols}
\bigskip
\address{
Departamento de F\'{\i}sica de la Materia Condensada, C-V\\
and Instituto de Ciencia de Materiales ``Nicol\'as Cabrera''\\
Universidad Aut\'onoma de Madrid, E-28049 Madrid, Spain}

\maketitle
\begin{abstract}
Supercurrent flow is studied in a structure that in the
Ginzburg-Landau regime can be described in terms of
an effective double barrier potential. In the limit of
strongly reflecting
barriers, the passage of Cooper pairs through such a
structure may be viewed as a realization of resonant
tunneling with a rigid wave function. For interbarrier
distances smaller than $d_0=\pi\xi(T)$ no current-carrying
solutions exist. For distances between $d_0$ and $2d_0$,
four solutions exist. The two symmetric solutions obey a
current-phase relation of $\sin(\Delta\varphi/2)$, while the
two asymmetric solutions satisfy $\Delta\varphi=\pi$ for all
allowed values of the current. As the distance exceeds $nd_0$,
a new group of four solutions appears, each contaning $(n-1)$
soliton-type oscillations between the barriers.
We prove the inexistence of a continuous crossover
between the physical solutions of the nonlinear
Ginzburg-Landau equation and those of the corresponding
linearized Schr\"odinger equation.
We also show that under certain conditions a repulsive delta
function barrier may quantitatively
describe a SNS structure.
We are thus able to predict that the critical current of a
SNSNS structure vanishes as $\sqrt{T'_c-T}$, where
$T'_c$ is lower than the bulk critical temperature.

\end{abstract}
\vspace{.5cm}


\vspace{.5cm}
\narrowtext

\section{Introduction}

The superconducting state is characterized by
the existence of
long-range phase coherence
in the electron system, and
its characteristic nondissipative currents are associated
to spatial
distortions of the macroscopic phase.
Superconductivity is however not
the only instance of electron transport being
phase-coherent
over distances much larger
than atomic length scales.
Another important example is provided by
mesoscopic normal transport, which began to be investigated
about fifteen years ago and which quickly matured
into an active branch of solid state physics.\cite{imry86}
Mesoscopic transport is realized
at low enough temperatures, when electrons
may preserve its phase coherence over long distances
and thus undergo multiple coherent scattering by impurities
or boundaries. Interest in this area stimulated experimental
and theoretical research on
a rich variety of mesoscopic phenomena
in solids, \cite{alth90} each of which is associated to
a specific quantum interference process. Let us consider
for a moment one of the most characteristic examples: the
Aharonov-Bohm (A-B) effect. This is the effect by which
the electronic properties of a thin ring depend periodically on the threaded
magnetic flux. As a mesoscopic phenomenon, it was first observed
in narrow normal cylinders \cite{shar81} and later
in small rings.\cite{webb86}
However, the A-B effect in solids had already been observed
in superconducting rings
not long after the discovery of the Josephson effect.\cite{jakl64}
This observation was possible
without the availability of modern nanotechnology
because of the existence of long-range phase coherence in
superconducting rings.
Phase rigidity of the superconductor collective wave function
is enforced by the spontaneous breaking of gauge symmetry.
In contrast, the spatial coherence of the electron field
in a normal system cannot rely on the existence of a phase
transition and requires
low temperatures and short length scales to make itself noticed.
One may adopt the simple picture that the A-B effect in a
superconducting ring relies on the
phase coherence of the Cooper pair
wave function, while its normal ring counterpart requires
coherence of the single electron propagator. This consideration
leads us to the question of whether,
for any given electron interference phenomenon
observed in a normal mesoscopic system, there may be
of a Cooper pair analogue that could eventually be observed
in a macroscopic superconductor. We have seen that there is an
A-B effect for single electrons as well as for Cooper
pairs. In the first case, the conductance (more generally,
the current-voltage characteristics) is a function
of the flux with a strong periodic component.\cite{webb86}
In the second case, it is the critical current (more
generally, the current-phase relation)
that depends periodically on the flux.
The question is whether other
quantum interference phenomena
may also display this dual nature whereby
they can be realized in the propagation of
single electrons or Cooper pairs.
A proper comparison of the two types of dynamics
requires the resolution of the equations satisfied
by the corresponding effective wave functions. In
the case of normal electrons, and within the noninteracting
approximation, one has to look at the wave function of
Fermi electrons. In the superconducting case, the
attention must be turned to the order parameter
$\psi({\bf r}) \propto <\hat{\psi}_{\uparrow}\hat{\psi}_{\downarrow}>$,
if one
wishes to obtain information on the global condensate
behavior.

Within the Ginzburg-Landau (GL) approximation,
$\psi({\bf r})$ must be an extremum
of the free-energy functional
\beq
F=\int d{\bf r}\{|(\nabla-i{\bf A})\psi|^2-
(1-V({\bf r}))|\psi|^2
+\lambda |\psi|^4/2\},
\label{glf}
\eeq
where $V({\bf r})$ is an effective potential and
reduced units have been used.\cite{sols94b,comm1} The parameter
$\lambda$ is introduced here to eventually
distinguish between the nonlinear
GL case ($\lambda\!=\!1$) and the linear case of a  Schr\"odinger
electron with reduced energy equal to one ($\lambda\!=\!0$).
Stationary solutions of Eq. (\ref{glf}) must satisfy the equation
\beq
(i\nabla+{\bf A})^2\psi - (1-V({\bf r}))\psi + \lambda |\psi|^2\psi = 0.
\label{gle}
\eeq
{}From the
study of Eq. (\ref{gle}), we plan to investigate
how Cooper pairs
behave under an effective potential $V({\bf r})$ that, in the absence
of nonlinearity, is known to yield a specific quantum interference
phenomenon.

Resonant tunneling (RT) is
one of the best understood and perhaps simplest
quantum interference phenomenona.\cite{chan91}
It can occur in one dimension without the requirement of a
nontrivial topology.
A quantum particle undergoes resonant tunneling when it
has to traverse a double barrier structure.
As a consequence of multiple inner reflection, the transmission
probability is very sensitive to such parameters as the
electron energy or the interbarrier distance, peaking near
well-defined resonances.
In this paper, we choose
resonant tunneling
as a case study in which to analyze the possibility
of a doublefold phenomenology in quantum interference
processes.
For simplicity, we focus on the case of a quasi--one-dimensional
superconductor and assume $\delta$-function barriers,
so that $V({\bf r})$ in Eq. (\ref{gle}) is taken to be of the form
\beq
V(x)=g[\delta(x-a)+\delta(x+a)].
\label{GLV}
\eeq

In the case of $g$ large and $\lambda=0$, Eqs. (\ref{gle}) and (\ref{GLV})
yield the essential phenomenology of RT for
independent electrons.
For this reason, we
will pay especial attention to the case $\lambda=1$ with
$g \gg 1$, although the
crossover to moderate values of $g$ in the nonlinear
case will also be
analysed.

We must specify
the transport properties we wish to compute. Ideally, one
would like to compare identical device properties, such
as the I--V curve, which largely characterizes a normal
RT diode. However, we already encounter at this stage
a major difference between the two types of systems we wish
to study. Unlike in the
case of normal transport, the I--V characteristics is not
the relevant quantity to obtain information about
the Cooper pair dynamics.
The application of an external bias $V$ creates a
nonequilibrium between the populations of quasiparticles
coming from each side of the structure. Only very indirectly
does this change in the quasiparticle population affect
the properties of the condensate, and the existing effect
is very sensitive to the dynamics of single
quasiparticles,\cite{sanc95} which is very similar to
that of normal electrons.
In contrast, the
population of Cooper pairs
cannot be driven out of equilibrium, since they
can only exist in a condensed state. One could at most
create a difference between the chemical potentials of
the condensates on each side of a structure. However,
in the case of two barriers, one
only expects a straightforward realization of the ac Josephson
effect.

The current carried by the condensate is driven by
spatial variations of the macroscopic phase. For this
reason, we
focus on the study of the current-phase relation
for configurations $\psi(x)$ satisfying Eq. (\ref{gle})
with $V(x)$ given by (\ref{GLV}).
The current-phase relation is only meaningful for a
superconducting structure; thus, we can hope at best to
give a qualitative comparison between the two types of
RT. One might still foster hopes that something resembling
the scattering wave function of a single electron might still
be obtained in the formal limit of $\lambda \rightarrow 0$.
As we will see, however, this is not case.
Because of the difference in boundary conditions,
the solutions for the superconducting
order parameter obtained from Eq. (\ref{gle}) with $\lambda\!=\!1$
and the wave functions
for Schr\"odinger scattering states with $\lambda\!=\!0$
fall into qualitatively
different classes of solutions that do not evolve continuosly
into one another as $\lambda \rightarrow 0$.

The physical explanation to the major differences between
the normal and superconducting versions of RT
is to be found in
the existence of
strong correlations among Cooper pairs, which is caused by their
bosonic nature and by their large mutual overlap.
The picture of independent particles undergoing
coherent multiple scattering, which often applies for Schr\"odinger electrons
in the normal state,
completely breaks down in the case of Cooper pairs. The notion
of many independent pairs which individually experience
quantum interference and collectively combine to form a macroscopic
wave function would at best be adequate for a condensate formed
by noninteracting molecular pairs. However, such a
scenario is not known to occur in nature. One may have
superfluids made of strongly interacting bosons (case of $^4$He)
or largely overlapping pairs ($^3$He), and superconductors
with large (BCS) or moderately short coherence length (case
of cuprate superconductors).
The strong correlation in conventional superconductors manifests
itself through the nonlinear term in Eq. (\ref{gle}).
As is characteristic of broken symmetry states, the
resulting wave function is rigid and does not obey a
superposition principle. We will see that the stiffness of the macroscopic
wave function is the primary cause of the important differences
between the Cooper pair and the individual electron scenarios.

Let us now return to the A-B effect. We have already noted
that it can be
observed both
in superconducting and normal (mesoscopic) rings.
Being a direct
consequence of electromagnetic gauge
invariance, \cite{bloc70}
the A-B effect is of a very fundamental nature.
The transport properties of a thin ring must always display
a periodic (although not necessarily
always observable) dependence on the
magnetic flux.
In contrast, the sensitivity of the current to, e.g.,
the interbarrier distance
(for a given
electron energy and for
a given normalization factor)
in a RT structure does not stem from any fundamental symmetry.
We will see that a rich dependence of the structure of
the set of solutions on the interbarrier distance does certainly exist.
We will also conclude however
that, at large separations, an important physical
quantity like the critical current becomes
essentially independent
of the interbarrier distance.

The search for a Cooper pair analogue of resonant tunneling has
led us to investigate the interesting properties
of superconducting structures that can be
described by a free-energy functional of the type (\ref{glf}). As is proved
in Appendix A, under certain conditions, a quasi--one-dimensional
SNS structure
without current concentration (S and N have the same width)
can be quantitatively described by a $\delta$-function in
the effective GL equation. As a consequence, the model given by Eqs.
(\ref{gle})
and (\ref{GLV}) can yield a quantitative description of a SNSNS structure
of uniform width. We have found that the set of
solutions to the corresponding GL equation can display
a very rich structure.
The main experimental prediction is that, in a SNSNS  structure,
the critical current vanishes with a law $I_c(T) \sim \sqrt{T'_c-T}$,
which differs markedly from the $(T_c-T)^2$ behavior of the
SNS case.\cite{abri88} In addition, we find $T'_c < T_c$, i.e.,
for a SNSNS system, there is a depression (with respect
to the SNS case) of the temperature at which the critical
current vanishes.

This paper is arranged as follows. After presenting the
model in section II, an analytical study of the solutions
to the nonlinear Eq. (\ref{GLmod}) is presented in section III.
In section IV, we prove the inexistence of a continuous
crossover between the physical solutions of
Eq. (\ref{gle}) and those of its linear version.
Some experimental predictions on temperature dependent
transport properties are discussed
in section V.
Finally, in section VI, we present some concluding remarks and
comment on how some of
the qualitative conclusions of the present article can
be extended to other interference phenomena.

\section{The model}

In this paper, we wish to analyse the
current-carrying solutions of the nonlinear GL equation
(\ref{gle}) with the effective potential given by (\ref{GLV}).
We may factorize the order parameter $\psi=R e^{i\varphi}$
and write ${\bf a = A - \nabla\varphi}$ for the superfluid velocity.
The gauge-invariant electric current is then written
${\bf j} = - R^2 {\bf a}$. Within the assumption
of a sufficiently narrow wire
(width much smaller than the coherence and penetration lengths)
it is
safe to neglect the dependence of ${\bf j}$
and $\psi$ on the transverse variables.
In these
conditions, our analysis reduces to the study of the solutions of the
nonlinear differential equation
\beq
\frac{d^2R}{dx^2}
+[1-g\delta(x-a)-g\delta(x+a)]R -\frac{j^2}{R^3}-\lambda R^3=0.
\label{GLmod}
\eeq
for arbitrary values of $g$ (hereafter, $\lambda\!=\!1$, unless
otherwise stated).
In a quasi--one-dimensional superconductor, we may choose
${\bf A}=0$
and write for the phase
\beq
\varphi(x) = j \int_{0}^{x} \frac{dx'}{R^2(x')}\;\;,
\label{PhDef}
\eeq
where the current density $j$ is
a conserved number, the total current through a lead of cross-section
$A$ being $I=jA$. For the superconducting order parameter we
are interested in solutions satisfying the boundary conditions
\beq
\left.\begin{array}{l}
R=R_{\infty}\\
\varphi(x) = qx \pm \Delta\varphi/2
\end{array}\right\}\;\;\;\mbox{for}\; x\rightarrow \pm \infty,
\label{BCond}
\eeq
$R_\infty$ being the biggest solution (i.e., that with the
lowest free energy) of the equation
\beq
R^6-R^4+j^2=0.
\eeq
In Eq. (\ref{BCond}), a nonzero value of $q=j/R^2_{\infty}$ accounts for
a linear variation of the phase at infinity,\cite{sols94a} and
the possibility of a phase offset
\beq
\Delta \varphi \equiv \int_{-\infty}^{\infty} (\varphi'(x)-q) dx
=j \int_{-\infty}^{\infty} (\frac{1}{R^2(x)}-\frac{1}{R^2_\infty}) dx
\label{OffDef}
\eeq
has been introduced.
As in the single barrier case,\cite{sols94a}
$\Delta\varphi$ will turn out to be
a most convenient parameter to classify
the solutions of Eqs. (\ref{GLmod}) and (\ref{BCond}).

One may wonder whether, apart from describing the condensate
analogue of resonant tunneling, the double delta barrier model
may quantitatively describe a specific physical system. Fortunately,
the answer is yes. It is proved in Appendix A that,
within the GL
approximation, a $\delta$-function effective barrier can
serve as a quantitative model
for a SNS structure.
More specifically, we show that,
if $\xi_N$ is the coherence length of the normal
metal, the effect of a normal segment of length $L$
inserted in
a superconducting wire can
be rigorously modeled by a $\delta$-function of strength
\beq
g=\frac{L\xi(T)}{\xi^2_N},
\label{GDef}
\eeq
provided that $L\ll \xi_N \ll \xi(T)$, where $\xi(T)$ is the
temperature dependent coherence length.
Since this effective delta barrier yields the correct matching
properties of a realistic SNS system,\cite{volk74} we may conclude
with confidence
that a model with a double $\delta$-function barrier will correctly
describe a SNSNS system.\cite{comm2} In addition, the fact that both
the barrier strength $g$ and the effective length $a$ are
functions of temperature for a given physical structure
(note that $a=d/2\xi(T)$, where $d$ is the physical distance
between the normal islands) permits us to use temperature
as a convenient driving parameter to
tune $g(T)$ and $a(T)$. This property has
important experimental consequences that will be
discussed in section IV.

Regarding the temperature dependence of the effective parameters,
it is interesting to note that, as $T \rightarrow
T_c$, one has $g(T) \rightarrow \infty$ and $a(T) \rightarrow 0$.
Therefore, the Josephson limit ($g \gg 1$) can always be explored
by driving $T$ sufficiently close to $T_c$.
In the Josephson
regime, the currents are necessarily much smaller than the bulk
critical current. As a consequence, the spatial variation of the
phase can be safely neglected for many purposes and the
phase offset can be identified with the
the conventional ``phase difference'' between the two superconducting
terminals,
$\Delta\varphi \simeq \varphi_1 - \varphi_2$.
In Refs. \cite{sols94a,lt20}
it was shown that, for large $g$, the critical current in the
presence of a $\delta$-barrier is
\beq
j^{(1)}_{c}=1/2g.
\label{JcSNS}
\eeq
Since the bulk
($g=0$) critical current is, in these units, $j_b=0.385$, one
can derive the relation
\beq
g=1.30\cdot(j_b/j^{(1)}_{c}),
\eeq
valid for large $g$. Going beyond the $\delta$-barrier model,
one can show that the critical current for a
SNS system of arbitrary normal length L is \cite{jaco65}
\beq
j_c^{(1)}=\left[\frac{2 \xi(T)}{\xi_N} \sinh(\frac{L}{\xi_N})\right]
^{-1},
\label{JCbarrier}
\eeq
provided that the resulting $j_c$ is much smaller than $j_b$.
One may readily note that the result (\ref{JCbarrier}) is consistent with
Eqs.~(\ref{GDef}-\ref{JcSNS}) in the limit $L\ll \xi_N$.

\section{Exact solutions}

\subsection{General properties}

In this section, we study the solutions
of Eq. (\ref{GLmod}) satisfying the boundary conditions (\ref{BCond}).
There is a mechanical analogue that
helps to understand some general properties of its
solutions.\cite{lang67} Eq. (\ref{GLmod}) may be viewed as the
force equation for a classical particle of unit mass
with position $R$ at time $x$ moving under a potential
\beq
v(R)=-\lambda R^4/4+R^2/2+j^2/2R^2.
\label{veff}
\eeq
In the GL case ($\lambda\!=\!1$),
configurations represented
by points A and B of Fig. 1a satisfy the asymptotic condition $dR/dx=0$,
with B energetically more favourable.
The picture is that of a classical particle which, at remote times
($x \rightarrow -\infty$),
stays in point B. At negative times, it
begins to roll down and, after receiving two kicks at times
$x= \pm a$, it
returns asymptotically ($x \rightarrow \infty$)
to point B. At each pulse, the change in the mechanical
energy $\varepsilon \equiv R'^2/2+v(R)$ is determined by the matching
condition
\beq
R'(x^+)-R'(x^-)=g R(x),
\label{DMatch}
\eeq
at points $x = \pm a$. Fig. 1b shows one particular solution $R(x)$
that exactly correlates with the mechanical analogue schematically
depicted in Fig. 1a.

The energy before and after the two kicks is $\varepsilon=v(R_\infty)$.
Eq. (\ref{DMatch}) indicates that the effect of barriers is that of
making the velocity $R'(x)$ more positive. If the system
is to return to point B at $x\rightarrow\infty$, it is not difficult to
see that the
conditions $\delta \varepsilon_0<0$ (where $\delta \varepsilon_0$ is the
change in mechanical energy at point $x=-a$)
and $R'(-a^-)<0$ must be satisfied, and, for
analogous reasons, $R'(a^+)>0$. These two constraints
force the solutions to be of the form \cite{abram}
\beqa
R^2(x)=Z + U \tanh^2(k(x-x_{+(-)})) & \; \; \; x>a \; (x<-a)
\nonumber \\
R^2(x)=e_1 + (e_2-e_1) \,\mbox{sn}^2 \left( \left.\sigma
\sqrt{\frac{e_3-e_1}{2}}
(x+a)+F_0 \right|
\frac{e_2-e_1}{e_3-e_1}\right) & \; \; \; |x|<a
\label{GLSol}
\eeqa
where $Z$ is the smallest root of $Z(Z-2)^2=8j^2$, $U\equiv 1-3Z/2$,
$k\equiv\sqrt{U/2}$, and sn$(x|m)$ is a Jacobi elliptic
function.\cite{abram}
Defining the function $y(x)\equiv R^2(x)$ and the parameters
$y_0\equiv y(-a)$ and
$y'_0\equiv y'(-a^-)$ (both of them
functions of $x_-$), the matching condition (\ref{DMatch})
at $x=-a$ determines $e_1, e_2, e_3$ as the roots of the polinomial
\beq
P(s)\equiv (s-Z)(s-1+Z/2)^2+4 \delta \varepsilon_0 \; s ,
\eeq
with $e_1 \leq e_2 \leq e_3$, and where the change in $\varepsilon$
is
\beq
\delta \varepsilon _0= g (y'_0+g y_0)/2.
\label{VarE}
\eeq
Finally,
\beq
F_0 \equiv F \left( \left. \arcsin \sqrt{\frac{y_0-e_1}{e_2-e_1}}
\right|
\frac{e_2-e_1}{e_3-e_1}\right)
\eeq
where $F(\varphi|m)$ is
the incomplete elliptic integral of the first kind,
\cite{abram} and $\sigma$ in (\ref{GLSol}) is defined as
\beq
\sigma \equiv \mbox{sgn}(y'_0 + 2 g y_0).
\eeq
In the limit $(y'_0 + 2 g y_0) \rightarrow 0$, it can be shown that both
values of $\sigma$ in Eq. (\ref{GLSol}) lead
to the same result for $R^2(x)$.

In analogy with $y_0$ and $y'_0$, we define $y_1\equiv y(a)$
and $y'_1\equiv y'(a^-)$. The second kick will take us
asymptoticaly to the value $R_\infty$ if it causes an energy change
$\delta\varepsilon_1$ such that
$\delta\varepsilon_1 + \delta\varepsilon_0 = 0$.
This leads us to write the relationship
\beq
y'_1+g y_1 + y'_0+g y_0=0
\label{Matching}
\eeq
as the global matching condition.
Eq. (\ref{Matching}) determines implicitly
the parameter $x_-$ to be introduced
in Eq. (\ref{GLSol}). Through the
integration of Eqs. (\ref{GLmod}-\ref{PhDef}),
each possible value of $x_-$ uniquely determines
one solution of Eq. (\ref{GLmod})
satisfying
the boundary conditions (\ref{BCond}).
Thus, $x_-$ is
a parameter that completely
characterizes a given physical solution.
Our goal is therefore to
solve numerically for all possible values of $x_-$ satisfying
Eq. (\ref{Matching}) for a given value of $j$.

A quantity of interest is the critical current, which we plot
in Fig. 2 as a function
of the (reduced)
semi-distance between barriers for several values of the barrier
strength. The result is given in
units of the critical
current for a single barrier with the same strength,
$J_c(a,g) \equiv I_c(a,g)/I_c^{(1)}(g)$.
For separation distances much larger than $\xi(T)$ the critical
current becomes identical to that of a single barrier, regardless
of the value of $g$. This is the limit in which the two barriers
are effectively decoupled. The decoupling at long distances is
caused by
the nonlinear term in Eq. (\ref{GLmod}). In contrast,
the dynamics of an electron obeying
the linear Schr\"odinger equation is always sensitive to the presence
of both barriers.
We will see however that, in the nonlinear case,
there are always solutions with high free energy
that are sensitive to the
presence of both barriers (in the sense that they cannot be viewed as
simple combinations of single barrier solutions), but these
energetic solutions become
increasingly irrelevant at large separations.

We observe in Fig. 2 that, for moderate values of $g$, there
is a slight depression of the critical current with respect to the
single barrier case, if the reduced semidistance is $a \alt 1$
(a similar effect was noticed in Ref. \cite{way77}
for the case of weak links). This
effect becomes more pronounced
as $g$ gets larger. For $g=10$, the law $1/2g$ already applies
approximately for
the critical current of single barrier and then
$J_c \simeq 1/2$ for $a=0$, as expected from a single barrier with
doubled strength. Deep in the Josephson regime
($g=20$ or larger), the interval $0<a<\pi/2$ is practically depleted
of solutions and the critical current is essentially zero. As will be seen
later,
this has noticeable experimental consequences, since the value of
the reduced distance is temperature dependent. In the following
two subsections,
we analyse the main properties of the solutions of of Eqs.
(\ref{GLmod}-\ref{BCond})
over the whole range of possible values of $g$.

\subsection{Josephson limit ($g \rightarrow \infty$)}

In the limit of large scattering strength ($g \rightarrow \infty$),
the presence of a single barrier is known to yield the ideal
Josephson behavior \cite{sols94a,lt20}
\beq
j=(1/2g)\sin(\Delta\varphi)
\label{JRel},
\eeq
where the phase offset $\Delta\varphi$ can be identified with
the conventional ``phase difference'' between the two weakly
linked superconductors.
For the double barrier case,
the parameter $\delta\varepsilon\equiv -\delta\varepsilon_0$
serves to characterize the continuous set
of solutions in the regime of large $g$.
For a given value of the current $j$, it is
shown in Appendix B that $\delta\varepsilon$
takes values
between 0 and $(1-J^2)/4$, where $J\equiv 2gj$ is the current
in units of the critical current for a single barrier structure.
We also show in Appendix B that the matching condition (\ref{Matching}) can be
rewritten as
\beqa
a=n \, G(\delta\varepsilon), \;\;\;\; n=1,2,...
\nonumber \\
G(\delta\varepsilon) \equiv
\left(\frac{2}{1+2\sqrt{\delta \varepsilon}}\right)^{1/2}
\, K\left(\frac{1-2\sqrt{\delta \varepsilon}}
{1+2\sqrt{\delta \varepsilon}}\right),
\label{JMatch}
\eeqa
where $K(m) \equiv F(\frac{\pi}{2}|m)$ is
the complete elliptic integral of the first kind.\cite{abram}
For each integer $n$ that meets the requirement
(\ref{JMatch}) there are four solutions $R(x)$,
all of them corresponding to the same
value of $\delta\varepsilon$. The quantity
$2G(\delta\varepsilon)$
is the spatial period of the solution between barriers. Thus, for
each solution of Eq. (\ref{JMatch}) that we may find for a given value of
$n$ in a structure of interbarrier distance $2a$, we can always construct
a solution for the structure of distance $2(a+G)$ whose index is $n+1$
and which is identical to the previous one except for the presence of
an extra oscillation.

	The structure of the solutions can be clearly appreciated in Fig. 3.
In the Josephson limit, we see that there are no solutions for
$2a<\pi$ ($J_c=0$). For $2a>\pi$
the critical current becomes nonzero and a group of
four solutions appears, two symmetric and
two asymmetric under the inversion $x\rightarrow -x$,
the two asymmetric ones being the
mirror image of each other.
Their current-phase relation is shown
in Fig. 4a for $a=1.74$.
The symmetric solutions form two branches that combine to
yield the curve
\beq
I = I_c(a,g) \sin(\Delta\varphi/2),
\label{JSyOff}
\eeq
which is the naive expectation
for two Josephson junctions in series. In contrast, the two asymmetric
solutions obey the law
\beq
\Delta\varphi=\pi, \;\;\;\mbox{for all $I$}.
\label{JAsOff}
\eeq
This peculiar current-phase relation can
be simply understood if the two junctions are assumed to be sufficiently
far apart. In this case,
the central peak in $R(x)$
between the two barriers becomes a long plateau
and the two barriers
behave as independent. For a given current, there
are two possible values of the phase, $\Delta\varphi_+$ and
$\Delta\varphi_-$, satisfying Eq. (\ref{JRel}). In the presence of
the two barriers, these two phase values may combine in four different
ways. In two cases, the phase difference is the same in both barriers.
Then, $\Delta\varphi$ equals $2\Delta\varphi_{+}$ or
$2\Delta\varphi_{-}$. This results in a
$\sin(\Delta\varphi/2)$ law for the symmetric solutions.
In the other two cases, a different phase change takes place
in each barrier. But the total phase difference $\Delta\varphi$
is given by the sum of the two values, which is exactly
\beq
\Delta\varphi_+ + \Delta\varphi_- = \pi.
\eeq
This explains Eq. (\ref{JAsOff}) and
the double vertical branch shown in Fig. 4a. The above set of
considerations
makes the results (\ref{JSyOff}) and (\ref{JAsOff}) quite plausible,
and even expected, for the case of large separations.
The not so obvious result is that the same conclusions hold when the separation
distance is comparable to $\xi(T)$ and the two barriers can in no way
be viewed as decoupled. This result can be proved rigourously from
analytical considerations (see Appendix C).
The smallness of the distance between barriers
makes itself noticed only through a depression of the critical current
with respect to the single barrier case, but {\it not} in the qualitative
form of the $I(\Delta\varphi)$ curves.

As the distance grows, the maximum current of the group of four solutions
saturates to the critical value for a single barrier.
As indicated in Fig. 3, a new
group of four solutions appears when $a>\pi$, and their
current-phase relation is shown in Fig. 4b.
Comparison with Fig. 4a reveals two features of the
$I(\Delta\varphi)$ relation:
(i) for $a$ not much larger than $\pi$,
the maximum current value of the second group of solutions has not
yet reached the saturation value, and (ii) all the values of $\Delta\varphi$
are shifted by $\pi$ with respect to the first set of four solutions.
Again, these features are easy to understand if the two barriers are
assumed to be far apart. The new group of solutions resembles the first
set of four, except in that a soliton (i.e., a spontaneous local
depression of the gap \cite{sols94a}) contributing an extra phase
of $\pi$ has nucleated between the barriers. As the two barriers
separate, the added depression of the order paramater, which
may initially be viewed as the result of quantum interference oscillations,
evolves into a well-defined, isolated soliton. Remarkably,
the additional depression in $|\psi|$ contributes exactly $\pi$ to the
phase, even when the two barriers are relatively close and the
depression does not act as an isolated kink but rather as an additional
oscillation. It must also be noted that, since $j\ne 0$, we have
$|\psi| \ne 0$ at all points.

As the separation distance is made increasingly larger, the
same pattern repeats itself. Each time the distance
$2a$ exceeds an integer multiple of $\pi$, a new group
of four solutions appears, with a structure similar
to that of the preceding set of four solutions
except for an extra phase of $\pi$.
As the
interbarrier distance continues to increase,
new sets of solutions
emerge periodically, always in groups of four.
For a given distance,
the groups of solutions that are most sensitive to the
double barrier feature (as indicated by its not yet saturated
maximum value) are those with a higher free energy resulting
from a higher number of modulations in the order parameter.
These very energetic solutions can be expected to be
irrelevant in practice, except perhaps to account for fine details
in the dynamic behavior. Therefore, it may be stated with reasonable
confidence that the two barriers become effectively decoupled
for practical purposes (also in regard to the dynamic behavior) when
the maximum current of the second group of solutions has reached
the saturation value. This happens approximately for $2a>3\pi$.

{\it Computation of the critical current.}
Eq. (\ref{JMatch}) (see also Fig. 3) indicates that $\delta\varepsilon$
is a unique function of
$a$ and $n$, so that the product in the r.h.s. hand side of Eq.
(\ref{JVarEn}) must be constant within
a given group of four solutions. The
highest possible value of $J$ (which is always to be found in the
$n=1$ group) is obtained by imposing
$\omega(1-\omega)$ to take its maximum value of 1/4. Thus we
conclude that the critical current $J_c\equiv 2gj_{c}$ is given
by the relation
\beq
a=\left(\frac{2}{1+\sqrt{1-J_c^2}}\right)^{1/2}
K \left(\frac{1-\sqrt{1-J_c^2}}{1+
\sqrt{1-J_c^2}}\right).
\label{JJc}
\eeq
Eq. (\ref{JJc}) yields the curve $J_c(a)$ for the critical
current in the Josephson
limit, and the result has been plotted in Fig. 2.

\subsection{Intermediate and small values of g}

In Fig. 5 one can analyze the crossover between the
regimes of large and small values of $g$ for the same
barrier distances we considered in the discussion of
the Josephson limit (see Fig. 4). Let us
comment the case of $a=1.74$. As we
depart from the limit of very large $g$, an extra branch
appears of solutions with $\pi \le \Delta\varphi \le 2\pi$ and
with a low critical value (see the graph in Fig. 5a for $g=20$).
As $g$ decreases further, the low current solutions with
$\Delta\varphi$ near $2\pi$ disappear and the resulting
branch begins to shrink until it merges (for $g \sim 0.2$)
with the solutions that formed the double vertical branch in
the Josephson limit. For
$g\rightarrow 0$, the two sets of solutions with $\Delta\varphi \ne 0$
form a triple branch
corresponding to three single solitons nucleated at points
$x_0=0$ and
\beq
x_0=\pm \frac{1}{k}\mbox{arctanh} \left( \frac{\sqrt{3 \alpha^2-1}}
{\alpha \sqrt{3-\alpha^2}}\right), \; \; \;
\alpha\equiv\tanh(ka)
\eeq
This result can be obtained analytically from the matching equation
(\ref{Matching})
and from the properties of the solutions (\ref{GLSol}) in the limit
$g \rightarrow 0$.
Meanwhile, in the same process of decreasing $g$, the branch of
solutions with the smallest
phase offset evolves towards the set of uniform solutions (those
for which $R(x)=R_{\infty}$
and $\Delta\varphi=0$)
that are characteristic of a perfect superconductor.\cite{sols94a}

As can be seen in Fig. 5b, the
crossover presents similar characteristics for the case where $a=3.5$.
Like in the Josephson limit, the main differences for large
$g$ (with respect to previous
shorter distance case) lie in the extra value of $\pi$ of
$\Delta\varphi$ shown by the second set of solutions and in the presence
of a four-fold branch with a saturated critical current value.
As $g$ decreases, the evolution of the four solutions with higher
phase offset runs similar to that of their counterparts in Fig. 5a.
New interesting features appear however for low values of $g$.
At $g=0.5$ and 0.2 it is clear that the two groups of four solutions
begin to merge into a simpler pattern of single and double
solitonic solutions (with $\Delta\varphi$ tending to $\pi$ and
$2\pi$ as $J\rightarrow 0$, respectively). Since double soliton
solutions cannot exist in the transparent case, the corresponding
(four-fold) branch begins to detach from the rest of curves and to
decrease its critical current until it eventually disappears.
For $g \rightarrow 0$ the remaining branch is identical to
its counterpart in the shorter distance case: three solutions
of single solitons located in the same points as above
and one uniform branch with zero phase offset.
Incidentally, it can be shown that for
$a \leq (2k)^{-1}
\ln(2+\sqrt{3})$ only the soliton at $x_0=0$ survives.
This is not the case however
in any of the two distances considered here.

An interesting consequence of studying the crossover between
large and small values of $g$ is that, among the curves
shown in Fig. 5, we can recognize similarities with
current-phase relations computed for other structures
that, not being exactly SNSNS, share some common features.
For example, Martin-Rodero {et al.} \cite{amr94}
have performed a self-consistent, zero-temperature
calculation of the
current-phase relation in mesoscopic weak links, modelled
by a tight binding chain linked to two broad Bethe lattices
that act as superconducting reservoirs. In spite of the obvious
differences between the two physical models, the similarity
between some of the resulting curves is striking. For
instance, the current-phase relation shown for the two
longest chains in Fig. 2c of Ref. \cite{amr94} resemble
some of the branches we obtain for $g=5$ and 20 in
the $a=1.74$ case. It is interesting to note that the
curves of Ref. \cite{amr94} that look alike correspond
precisely to the case of strong internal reflection
at the constriction.
One concludes: (i) the nonlinear term of the GL
equation has an effect very similar to that of self-consistency
in a zero-temperature calculation;
and (ii) the essential physics rests in the nonlinear effects
taking place within the finite superconducting segment and the
effective scattering at its two ends, the physical details
of the semi-infinite S leads being  less important. Conclusion
(i) is what one expects from inspecting the
microscopic derivation of the GL formalism, and it has
already been noted in connection with the crossover for
large to small $g$ in the single barrier case.\cite{sols94a}

\section{Inexistence of a continuous crossover between the
linear and nonlinear problems}

It has already been commented in the Introduction that
there are qualitative differences in the physics described
by Eqs. (2-4)
in the linear (Schr\"odinger) and nonlinear (GL)
cases. However,
one might still think that, from a mathematical point of
view, there could be a continuity (as a function of $\lambda$)
between the order parameter configurations satisfying the
GL equation ($\lambda=1$) and the scattering wave function for
a Schr\"odinger electron in the presence of potential (\ref{GLV})
with $\lambda=0$ and unit energy.
Again, it is useful to consider the mechanical analog represented
in Fig. 1. The wave function of a retarded scattering state
is of the type $\psi \sim e^{ikx}+r e^{-ikx}$, for $x\le -a$,
if the electron is coming from the left, so that
the amplitude $R=|\psi|$ displays oscillations in the
left asympotic region. The mechanical picture to introduce in Fig. 1
would be that of a particle which at remote times $x\rightarrow
-\infty$ is oscillating between two return points with energy
$\varepsilon < v(R_{\infty})$, receives two sudden pulses
at times $x=\pm a$, and ends asymptotically in one of
the stationary points (this corresponds to the
uniform amplitude solution $\psi \sim te^{ikx}$ to be found
in the $x\!>\!a$ region). Thus we see that a major difference
between the two cases
lies in the asympotic behavior of $R(x)$, which
is always uniform for the superconducting order parameter
while it may oscillate for
a Schr\"odinger electron. The only remaining possibility
to find a common mathematical wave function must be looked
for in that combination of parameters which yields a transmission
unity for the electron linear wave, since in this case $R$ would
be uniform on both sides of the structure. The equivalent
mechanical particle would start from a stationary point to
which it would return at late times. The problem is that, for
an electron satisfying the linear Schr\"odinger equation,
that stationary point has to be A (see Fig. 1), and not B,
as is the case for the GL order parameter. Although a rigorous
demonstration of this statement is possible, it suffices to note
that, as $\lambda \rightarrow 0$, the potential maximum at B
moves toward infinite values of $v$ and $R$ (see B' in Fig. 1
for $\lambda=0.8$), effectively disappearing
from the picture when $\lambda=0$. Thus the only stationary
point that is available in the linear case to describe purely
transmitted waves is precisely point A, qualitatively different
from B even in the limit $\lambda \rightarrow 0$. We conclude
that, as the nonlinear term in Eq. (\ref{GLmod}) is made to vanish
$(\lambda \rightarrow 0$), there
does not exist a continuous crossover between the
physical solutions of the nonlinear Ginzburg-Landau equation
and those of the linear Schr\"odinger equation.

\section{Experimental predictions}

The proximity between barriers is most noticeable when the low
energy solutions are sensitive to it. Of course, the most dramatic
effects can be seen for $2a<\pi$, when no solutions exist at all
in the large $g$ limit.
This has interesting experimental consequences.
We must first remember that
$a$ is the reduced semidistance in units of $\xi(T)$, i.e.,
$a(T)=d/2\xi(T)$.
This means that, for a given structure
with a fixed physical distance $d$, the reduced distance $2a$ can be
made arbitrarily small by driving the temperature sufficiently close
to $T_c$, since then $\xi(T) = \xi'_0(1-T/T_c)^{-1/2}
\rightarrow \infty$ (for a clean superconductor, $\xi'_0=0.74 \xi_0$,
while for a dirty
superconductor $\xi'_0=0.85 \sqrt{l \xi_0}$,  $l$ and $\xi_0$ being
the mean free
path and the zero-temperature coherence length, respectively \cite{tink}).
If the double
barrier structure is formed by two normal segments, we know
from the analysis in section III that $g(T)$ scales towards
the Josephson limit.
Therefore, there is
a temperature $T'_c < T_c$ above which $d<\pi\xi(T)$.
Since, at the same time,
$g(T)$ is very large, this has the consequence that
no current-carrying superconducting solutions are allowed in the
temperature interval $T'_c < T < T_c$. The result is that, for a
SNSNS structure, there is
a depression (with respect to the SNS case) in the critical temperature
above which the critical current becomes zero.

In Fig. 6,
the critical current
is plotted as a function of temperature for both
a SNS and a SNSNS structure.
The main feature of the resulting
$I_c(T)$, namely, the law with which it vanishes, is
amenable to an analytical treatment.
In section III, we
derived Eq. (\ref{JJc}), which determines
$J_c(T)$ for a given value of $a(T)$. Since we are
interested in finding out how the critical current
vanishes, we may expand the r.h.s. of Eq. (\ref{JJc})
for small values of $J_c(T)$ and obtain
\beq
a(T) \equiv \frac{d}{2\xi'_0}\sqrt{\frac{T_c-T}{T_c}} \simeq \frac{\pi}{2}
\left(
1+\frac{3}{16} J^{2}_{c}(T)\right).
\eeq
The temperature $T'_c$ for which $J_c(T'_c)=0$ is $T'_c=T_c-\Delta T_c$, with
\beq
\Delta T_c = \left(\frac{\pi\xi'_0}{d}\right)^2 T_c.
\label{dtc}
\eeq
Eq. (\ref{dtc}) is actually an upper bound to the
value of $T_c-T'_c$, and it tends to the exact value
when the condition $g(T'_c) \gg 1$ is consistently
satisfied.

After some simple algebra, we get for the critical
current (in real units):
\beq
I_c(T) =I_{c}^{(1)}(T'_c) (\frac{8}{3\pi^2})^{1/2}
(\frac{d}{\xi'_0})
\left(\frac{T'_c-T} {T_c}\right)^{1/2}
\;\;\;\;\; T\rightarrow T'_c,
\eeq
where
\beq
I_{c}^{(1)}(T)
=\frac{A}{g(T)} (\frac{e\hbar}{m}) \frac{\psi^{2}_{\infty}(T)}{\xi(T)}=
\frac{A}{L} (\frac{\hbar c^2}{16 \pi e}) (\frac{\xi_{N}}
{\kappa \xi'^{2}_{0}})^2 \frac{(T_c-T)^2}{T^{2}_{c}}
\eeq
is the critical current of a SNS
structure at temperature $T$, $A$ being the cross area of the junction
and $\kappa\equiv\lambda(T)/\xi(T)$ is the Ginzburg-Landau parameter of the
superconductor.
The square root behavior of $I_c(T)$ for a SNSNS structure
contrasts markedly
with the $(T_c - T)^2$ law of
the SNS case.\cite{abri88}

Simple estimates suggest that the predicted
depression in the critical temperature should be
measurable. For example, for a SNSNS structure made
with superconducting Al ($T_c=1.19$ K and $\xi_0=1.6$ $\mu$m)
with $d=10 \xi_0$
(which falls within the validity of GL regime)
one would obtain
a decrease of $\Delta T_c \sim 0.05$ K. On the other hand, it
is interesting to note that, if the
condition $d \gg \xi_0$ is satisfied, there is a range of temperatures
sufficiently far below $T_c$ for which $d\gg \xi(T)$ and, thus,
$I_c(T) \simeq I_c^{(1)}(T)$.

The sensitivity of the effective $g$ to temperature
can be exploited in other interesting ways. For example,
by varying the temperature, one may drive
a given SNS or SNSNS structure from the large to the small $g$
regime. Consider a S$\bar{\mbox{N}}$S structure made in
a narrow wire of finite length at whose extremes we
apply an external voltage. \=N
is a normal metal with critical temperature $\bar{T}_c < T_c$.
When $T$ gets close to $T_c$, $g(T)$ becomes large, and so does
$\xi(T)$.
Both facts contribute to yield an ideal Josephson behavior
(a large value of $\xi(T)$ makes the length of the wire effectively
shorter, thus facilitating the adiabatic response characteristic
of the ac
Josephson effect \cite{sols94a}).
As an external voltage is applied, the current oscillates very
rapidly and one observes a zero time-average,
$I_{\mbox{av}}=0$. As the temperature is lowered below $\bar{T}_c$, the
structure becomes of the type S\={S}S with an effective $g \alt 1$
(we may choose \=S to be not very different from S).
The situation is then close to that
of a uniform superconductor and the system cannot respond adiabatically to
the externally applied bias  \cite{sols94a,lang67,PMG}.
Some type of resistive behavior has to be displayed,
with the result that $I_{\mbox{av}} \ne 0$. The net effect is that,
by lowering the
temperature, it is possible to drive a given SNS structure
from adiabatic
to resistive response.

\section{Conclusions}

We have studied superconducting flow in structures which,
in the Ginzburg-Landau regime, can be described
by a double barrier effective potential.
We have found that the critical current is depressed
with respect to the single barrier case. In the limit of
strongly reflecting barriers, $I_c$ can become effectively
zero for interbarrier
distances $d<d_0\equiv \pi\xi(T)$.
As $d$ exceeds $d_0$, four solutions (two symmetric and two
antisymmetric) appear, as might be expected for two Josephson links
in series. As $d$ grows even larger, new solutions appear,
always in groups of four. The supression of $I_c$ for
short distances
has practical consequences, since $d_0(T) \rightarrow \infty$
as $T\rightarrow T_c$, and the condition $d\le d_0$ is always
reached by any device if $T$ is sufficiently close to $T_c$.
Thanks to this analysis,
we have found that the critical current of a
symmetric SNSNS structure is lower than that
of a SNS structure with an equivalent N segment.
More important, the law with which $I_c(T)$ vanishes
is qualitatively different
in each case (see Fig. 6): $I_c(T)$ decays as $\sqrt{T'_c -T}$ in
a SNSNS structure, as opposed to the $(T_c-T)^2$ behavior
of the SNS case, with the inequality $T'_c < T_c$
being always satisfied.

We have also shown that, because of the different boundary
conditions, the solutions of the GL equation cannot evolve
continuously towards the scattering solutions of the linear
Schr\"odinger equation as the nonlinear term is formally
made to vanish.

We close this article by returning to the question
that initially motivated it, namely, the possibility
of finding a macroscopic quantum analog of a
specific interference process such as the resonant tunneling
of Schr\"odinger electrons. It was already commented
in the Introduction that, given the important differences between
the relevant transport properties of the physical systems involved,
the analogy would at best be qualitative. We have learned
that a basic fact such as the dependence of transport
behavior on the interbarrier distance --all other parameters being
equal-- effectively disappears in a superconducting structure with
$d \gg \pi\xi(T)$ (see Fig. 2). Since $\xi(T)$ is the length
scale needed for the nonlinear term in Eq. (\ref{gle}) to make itself
noticed, we may adopt the view that the nonlinear term in the
GL equation acts as a dephasing term that damps any interference
effect requiring modulation of the wave function amplitude $R$
(as would be caused, i.e., by the interference of waves traveling
in different directions). This picture allows us to extrapolate
our conclusion on the inequivalence of macroscopic
and microscopic resonant tunneling to other interference phenomena.
Consider, for instance, weak localization. If one considers a
superconducting structure which, in the GL regime, is described
by Eq. (\ref{gle}) with a weakly disordered
effective potential $V({\bf r})$,
could we expect that quantum interference
might cause a reduction of, i.e., the critical current?
{}From the analysis presented here, the answer seems to be no,
at least when the distance between ``impurities'' is much
larger than $\xi(T)$: the nonlinear term would damp interference
effects to the point of making each impurity act as isolated.
In particular, one should not expect to observe the macroscopic
equivalent of Anderson localization.
A similar conclusion holds in principle for
other mesoscopic phenomena based on geometry-induced
quantum interference such as the modulation of current by
a tuning stub:\cite{sols89} one should not expect any
type of interference effect associated to the collective
wave function in a superconductor enclosed in a specific
(topologically trivial) geometry, as long as the relevant
length scales are of order
$d \gg \xi(T)$. As indicated in the Introduction,
real analogs are to be found only for those ``interference''
phenomena based on fundamental symmetries such as gauge
invariance, which gives rise to the
Aharonov-Bohm effect.
A qualitative conclusion is
that care must be taken when developing physical intuitions
based on the idea that
the superconducting order parameter plays the role of
a macroscopic quantum
wave function.

\acknowledgments

We wish to thank Jaime Ferrer for valuable discussions.
This work
has been supported by CICyT, Project no. MAT91-0905, by DGICyT,
Project no. PB93-1248, and by
the Human and Capital Mobility Programme of the EC.
One of us (I.Z.) gratefully acknowledges the support from the
Comunidad Aut\'onoma de Madrid through a F.P.I. fellowship.


\appendix
\section{Deduction of the barrier model for a SNS structure}

Suppose we are given a quasi-one dimensional SNS structure.
Let the normal metal ocuppy the region $[-L/2,L/2]$.
$T_{c}^{(N)}$ and $T_{c}^{(S)}$ are the critical temperatures of
N and S, respectively.
We have $T_{c}^{(N)}<T<T_{c}^{(S)}$ and assume $T_{c}^{(S)}-T_{c}^{(N)}
\ll T_{c}^{(N)}$.
Following Refs. \cite{abri88,zait65}, we assume
that both N and S have the same
order parameter to gap ratio, $\psi / \Delta$,
as well as the same quasiparticle mass.
It can be proved \cite{zait65} that,
in these conditions,
the Ginzburg-Landau equations are valid for all $x$, and that
the matching conditions are determined by the continuity of
$\psi$ and its first derivative at $x=\pm L/2$. Thus one has
\beqa
-\frac{\hbar^2}{2m}\frac{d^2R}{dx^2}+\alpha_SR+\beta_SR^3+
\frac{mj^2}{8e^2R^3}=0 & \; \; \; |x| > L/2 \\
-\frac{\hbar^2}{2m}\frac{d^2R}{dx^2}+\alpha_NR+\beta_NR^3+
\frac{mj^2}{8e^2R^3}=0 & \; \; \; |x|< L/2
\eeqa
where $R(x)=|\psi (x)|$ and $\alpha_N, \beta_N$ ($\alpha_S, \beta_S$)
refer to the Ginzburg-Landau parameters of the normal
(superconducting) metal,
all of which depend on temperature.

As $T$ approaches $T_c^{(S)}$, we can neglect the term
$\beta_NR^3$ in Eq. (A2) and the remaining parameters of the normal metal
can be replaced by their fixed values at the $T_c^{(S)}$.
Shifting to reduced units of the superconductor \cite{comm1} we write
\beqa
-\frac{d^2R}{dx^2}-R+R^3+\frac{j^2}{R^3}=0 & \; \; \; |x|>\l/2 \\
-\frac{d^2R}{dx^2}+\theta R+\frac{j^2}{R^3}=0
& \; \; \; |x|<l/2
\eeqa
where $l \equiv L / \xi (T)$. This model is identical to that
considered by Jacobson \cite{jaco65} with
\beq
\theta \equiv \frac{\xi^2(T)}{\xi^2_N} \gg 1.
\eeq
If $\theta \gg j_c^2/R^4$, the integration of (A4) along the normal
segment yields
\beq
R'(l/2)-R'(-l/2) \simeq l\theta R,
\eeq
where $\sqrt{\theta} l = L/ \xi_N \ll 1$ has been assumed, so that
$R(x)$ can be approximated as constant within the integral.
In these conditions, the effect of the normal metal can
be exactly mimicked by a delta barrier located at $x=0$ with a strength
\beq
g \equiv l \theta = \frac{\xi(T) L}{\xi^2_N},
\eeq
as we wished to prove.

\section{Matching in the Josephson limit}

If, from the parameters intervening in Eq. (\ref{GLSol}),
we define
\beq
z \equiv \tanh(k(a+x_-)),
\eeq
it is not difficult to see that $y_0$ and $y'_0$ defined in the text
are given by
\beq
y_0=Z+U z^2
\eeq
\beq
y'_0=\sqrt{2U^3}(z^3-z).
\eeq
So, the condition
$y'_0 \le 0$ implies $0 \le z \le 1$.
Eq. (\ref{VarE}) for $\delta\varepsilon_0$ can then be rewritten as
\beq
\delta \varepsilon_0(z)=\frac{g}{2} \left( \sqrt{2 U^3}(z^3-z)
+g(Z+U z^2) \right).
\label{PolVarE}
\eeq
The additional requirement
$\delta \varepsilon_0 < 0$ implies that $z$ is further
restricted to the range $z_{min} < z < z_{max}$, where
$z_{min},z_{max}$
are the roots of $\delta \varepsilon_0(z)$
lying between 0 and 1. We define $\omega$ as
\beq
z \equiv z_{min} + \omega (z_{max}-z_{min}),
\eeq
so that, obviously, $0 < \omega < 1$.

In the Josephson limit ($g\rightarrow \infty$) it is easy to see that
$Z \simeq J^2/2g^2$ ($J \equiv 2gj$) and $U \simeq 1$.
Since, in the same limit, the only surviving finite root
of $\delta \varepsilon_0(z)$
is $z=0$, we can neglect $z^3$ in front of $z$ in Eq. (\ref{PolVarE}).
Then $z_{min}$ and $z_{max}$ can be easily calculated as the roots
of
\beq
-\sqrt{2} z + \frac{J^2}{2g}+g z^2=0
\eeq
To leading order in $1/g$, we get
\beq
z_{max,min}=\frac{1}{g\sqrt{2}} \left(1\pm \sqrt{1-J^2} \right)
\eeq
and, from (B5),
\beq
z=  \frac{1+(2\omega
-1)\sqrt{1-J^2}}
{g\sqrt{2}} + o(\frac{1}{g}),
\eeq
where $o(1/g^n)$ stands for any expression such that
$\lim_{g\rightarrow \infty} g^n \; o(1/g^n) = 0$.
{}From Eq. (B4), we obtain, also to leading order,
\beq
\delta \varepsilon_0(z)=\frac{g}{2} \left( \sqrt{2} z
+\frac{J^2}{2g}+g z^2 \right)
\eeq
which can be easily shown to lead to
\beq
\delta \varepsilon \equiv - \delta \varepsilon_0= \omega (1-\omega)(1-J^2)
\label{JVarEn}
\eeq
For the other quantities in Eq. (\ref{GLSol}), we obtain
\beqa
e_1=\frac{J^2}{2(1-4\delta \varepsilon)} \frac{1}{g^2}+o(\frac{1}{g^2})
\nonumber \\
e_2=1-2\sqrt{\delta \varepsilon}+o(1)\nonumber \\
e_3=1+2\sqrt{\delta \varepsilon}+o(1)
\label{EJDef}
\eeqa
The quantities $\beta$, $\Omega_0$, $\Omega_1$, and
$m$ are defined as:
\beqa
F_0 \equiv\frac{\beta}{g}+o(\frac{1}{g}) \\
2 a \left(\frac{e_3-e_1}{2}\right)^{1/2} \equiv \Omega_0+\frac{\Omega_1}{g}
+o(\frac{1}{g}) \label{OmegaDef}\\
m \equiv \frac{e_2-e_1}{e_3-e_1}=\frac{1-2\sqrt{\delta \varepsilon}}
{1+2\sqrt{\delta \varepsilon}}+o(1)
\eeqa
In this limit, the parameter $\sigma$ of Eq. (\ref{GLSol}) becomes $+1$
and the global matching equation
(\ref{Matching}) can be rewritten as
\beq
y'(-a^+)+y'(a^-)=g(y(-a)-y(a)).
\eeq
Solving this equation order by order in $1/g$,
we get:
\beq
\mbox{sn}^2(\Omega_0|m)=0,
\eeq
which is equivalent to
\beq
\Omega_0=2 n \,
K\left(\frac{1-2\sqrt{\delta \varepsilon}}{1+2\sqrt{\delta \varepsilon}}\right)
\;\;\; n=1,2,...
\label{JOmega0}
\eeq
{}From Eqs.(\ref{EJDef}) and (\ref{OmegaDef}) we deduce
\beq
\Omega_0 \simeq 2 a \left(\frac{1+2\sqrt{\delta\varepsilon}}{2}\right)^{1/2}
\eeq
for $g$ large, and find that Eq. (\ref{JOmega0}) leads to Eq.
(\ref{JMatch}) in the text, which we wished to prove.

We also find that $\Omega_1$ can take values
$\Omega_1=-2\beta$ (for the two symmetric solutions) or
$\Omega_1=-\Omega_0/a$ (for the two asymmetric ones).
Solutions come in pairs because of the quadratic character
of Eq. (\ref{JVarEn}). The behavior under the transformation
$x \rightarrow -x$ can be deduced from the expression for
the solution, which to leading order in $1/g$
takes the form
\beq
y_n(x)\simeq e_1+e_2 \,\mbox{sn}^2\left(
\left.\frac{2 n K+\Omega_1/g}{2a}x+n K +
\frac{2\beta + \Omega_1}{2g} \right| \frac{1-2\sqrt{\delta \varepsilon}}
{1+2\sqrt{\delta \varepsilon}}\right)
\eeq
where $K$ stands for the elliptic integral in (\ref{JOmega0}).

\section{Offset in the Josephson limit}

One can insert the asymptotic expressions obtained for $R^2(x)$ in the
previous Appendix into the second integral of Eq. (\ref{OffDef})
and make use of the identity:
\beq
\lim_{g\rightarrow \infty} g \int_{N K(m) + \alpha /g}^{M K(m) -\beta /g}
\frac{du}{1+g^2 \gamma \mbox{sn}^2(u|m)}= \nonumber
\eeq
\beq
\frac{1}{\sqrt{\gamma}}\left( (\frac{\pi}{2}-
\arctan (\alpha \sqrt{\gamma}))\delta_N  + (\frac{\pi}{2}-
\arctan (\beta \sqrt{\gamma}))\delta_M  + \pi N_e \right),
\label{JInt}
\eeq
valid for integers $N,M$ and arbitrary positive
values of $\alpha,\beta,\gamma$, with $0<m<1$.
In Eq. (\ref{JInt}), $N_e$ is the number of even integers $p$
satisfying $N<p<M$, and $\delta_N$ is defined as one(zero) for $N$
even(odd).
After some lengthy but straightforward algebra, one can prove
that the current-phase relation for the solutions of
Eq. (\ref{JMatch}) is ($J>0$)
\beq
J(\Delta\varphi)=\sqrt{1-4\delta \varepsilon} \left| \sin \left(
\frac{(n+1)\pi-\Delta \varphi}{2}
\right) \right|, \; \; \; \; \Delta \varphi \in
[(n-1)\pi, (n+1)\pi],
\eeq
for the symmetric solutions, and
\beq
\Delta \varphi=n\pi, \;\;\;\mbox{for all $J$},
\eeq
for the asymmetric solutions.

\begin{figure}
\caption{
(a) Potential energy $v(R)$ (see Eq. (13) in the text)
for the mechanical analog of
Eq. (4) with $j=0.01$, $g=2$ and $a=2$,
for $\lambda=1$ (solid line) and
$\lambda=0.8$ (dotted line); the straight lines depict
one possible trajectory of the equivalent particle
that begins and ends at point B, the numbers in the arrows
indicating time-ordered flights between kicks (the distance
between the two horizontal lines is exactly $\delta\varepsilon$).
(b) Inverted representation of the
particular solution $R(x)$ of Eq. (4) whose equivalent
mechanical trajectory is that depicted in (a).
}
\end{figure}

\begin{figure}
\caption{
Critical current as function of the semidistance between barriers
for several values of the parameter $g$. The distance is given in
units of $\xi(T)$ and the current is given in units of the critical
current for a single barrier with the same $g$ (see text).
}
\end{figure}

\begin{figure}
\caption{
Representation of the matching equation (22) in the text.
The quantity $n G(\delta \varepsilon)/\pi$ is plotted for several
values of the index $n$. A set of four solutions exists for
each combination of $n$ and $\delta\varepsilon$ that meets
the requirement $nG(\delta\varepsilon)=a$, where $a$ is the
reduced semidistance between barriers.}
\end{figure}

\begin{figure}
\caption{
Current as a function of the total
phase difference for $a=1.74$ (a)
and $a=3.5$ (b), in the limit of
$g$ very large.
The current is given in units of the critical
current for a single barrier with the same $g$,
so that $\lim_{g\rightarrow\infty}
2gj$ is plotted.
}
\end{figure}

\begin{figure}
\caption{
Current as a function of the
phase offset for several values of $g$ in the cases $a=1.74$ (a)
and $a=3.5$ (b), in the same units as in Fig. 2.
}
\end{figure}

\begin{figure}
\caption{
Temperature dependence of the critical current
in the vicinity of the critical temperature, for
SNS and SNSNS structures without current concentration.
$d/\xi'_0=10$ has been taken, and units are such that the
the prefactor of $(1-T/T_c)^2$ in Eq. (31) equals unity.
}
\end{figure}


\begin{references}
\bibitem{comm1}
In these units, $\xi(T)$ is the unit of length,
the order parameter $\psi$ is measured in units of $\psi_{\infty}$
(absolute value of the bulk order parameter at zero current), and $2e(\hbar/m)
(\psi_{\infty}^2/\xi(T))=c\sqrt{2}H_c(T)/4\pi \lambda (T)$
is the unit for current ($H_c$ and $\lambda$ are the critical
magnetic field and the penetration depth, respectively).
\bibitem{comm2} Eq. (23) of Ref. 12 would only apply for $L \gg \xi_N$,
a limit for which, unlike stated there, we have not found a rigorous
justification for the use of a $\delta$-function model.
\end{references}
\end{document}